\documentclass[twocolumn,showpacs,preprintnumbers,amsmath,amssymb]{revtex4}
\usepackage{graphicx}
\usepackage{dcolumn}
\usepackage{bm}

\begin{document}

\title{Evolution of Fluctuation in relativistic heavy-ion collisions}

\author{Bedangadas Mohanty, Jan-e Alam and Tapan K. Nayak }

\medskip

\affiliation{Variable Energy Cyclotron Centre, Calcutta
   700064, India}

\date{\today}

\begin{abstract}

We have studied the time evolution of the fluctuations in 
the net baryon number for different initial conditions and space
time evolution scenarios. We observe that the fluctuations 
at the freeze-out depend crucially on the equation
of state (EOS) of the system and for realistic EOS 
the initial fluctuation is substantially dissipated
at the freeze-out stage. 
At SPS energies the fluctuations in net baryon number at the freeze-out 
stage for quark gluon plasma and hadronic initial state 
is close to the Poissonian noise for ideal as well as for EOS 
obtained by including heavier hadronic degrees
of freedom. For EOS obtained 
from the parametrization of lattice QCD results the fluctuation is larger than 
Poissonian noise.  It is also observed that
at RHIC energies the fluctuations at the freeze-out point deviates
from the Poissonian noise for ideal as well as realistic equation of state,
indicating presence of dynamical fluctuations.

\end{abstract}

\pacs{25.75.-q,05.40.-a,12.38.Mh}
\maketitle

\section{INTRODUCTION}

Heavy-ion collisions at relativistic energies offer a unique environment for the 
creation and study of the Quark-Gluon Plasma (QGP) phase of nuclear matter.
In the collision process the system may
go through a phase transition whereby it 
evolves from a very hot and dense QGP state to normal hadronic matter.
A characteristic feature of this process is that the system experiences large 
event-by-event fluctuations in thermodynamic quantities such as 
temperature,  number density etc. 
Our ability to observe a characteristic fluctuation in 
various observables, in present day experiments, has been facilitated by   
the large number of particles produced in the relativistic
heavy-ion collisions at the Super Proton Synchrotron (SPS) 
and the Relativistic Heavy-Ion Collider (RHIC) along with 
the advent of large acceptance detectors at these experiments~\cite{qm01}.
The near Gaussian distributions of the experimental
observables like multiplicity, transverse energy ~\cite{wa98_fluc},
transverse momentum and particle ratios~\cite{na49_fluc} have provided 
an opportunity to relate these quantities to thermodynamical properties 
of matter, such as specific heat~\cite{cv_prl} and matter compressibility
~\cite{compress}. Fluctuations as a signature of quark-hadron phase 
transition has been experimentally studied extensively but no definite 
signature has  
been observed in this sector so far~\cite{phenix,na49_fluc,wa98_fluc,henning}. 

The main task for using fluctuation as a probe of QGP phase
transition would be to identify an observable whose fluctuation 
will survive from the time of 
formation of plasma till they get detected after the freeze-out. 
In this context, 
it was suggested that the study of the fluctuation in conserved quantities 
{\it e.g.} net baryon number, net charge or net strangeness will be 
very useful~\cite{evs,twin_prl1,twin_prl2,prakash}. The fluctuation 
in any conserved quantity ($\cal O$) is given by ~\cite{landau}
\begin{equation}
(\Delta{\cal O})^2 \equiv 
\langle {\cal O}^2 \rangle - \langle{\cal O}\rangle^2 =
T\, {\partial\langle{\cal O}\rangle\over\partial\mu}\,
\label{eq1}
\end{equation}
where, $\mu$ is the associated chemical potential and $T$ is the temperature 
of the system under consideration. 
Here $\langle{\cal O}\rangle$ 
represents the mean value of the conserved quantity over a large number of 
events and $(\Delta{\cal O})$ denotes the deviation of its value from
the mean, event-by-event. It has been shown that the fluctuation in these 
conserved quantities with and without QGP formation are very different. 
Hence, study of fluctuations in these conserved quantities can be a good 
signature of transition from confined to de-confined state of matter.

However, in the earlier calculations~\cite{twin_prl1,twin_prl2}
ideal gas EOS for both the QGP and hadronic gas was considered. 
For the QGP initial state, the evolution of fluctuation 
during mixed phase have also not been taken into account.
Finally, modifications of hadronic spectral functions in the thermal
bath was ignored.

The aim of this paper is to study the evolution of the fluctuation in conserved 
 quantities from time of formation to the time of detection. In addition to the
 QGP and hadronic initial state we consider an important case where the 
 hadrons are formed at the initial stage but the spectral function of 
the hadrons 
 are different from
 their vacuum properties. We also discuss the sensitivity of the results on
 different EOS. Here we concentrate only on the net baryon number 
 fluctuation and the results are presented in terms of the experimentally
 measured quantity, $(\Delta N_b)^2/N_b$, where $N_b$ is the number of baryons 
 and $\Delta N_b$ is the fluctuation in the net baryon number.

The paper is organized as follows. In the next section we
discuss the possible evolution scenarios. Then the initial conditions used 
for solving the evolution equations for various evolution scenarios are 
presented in section III. In section IV we study the
evolution of the system. Section V and VI is devoted 
for the results at the freeze-out for SPS and RHIC energies respectively. 
And finally in section VII we present the  summary and conclusions.

\section{EVOLUTION SCENARIOS}
We consider the following three possible evolution scenarios.

\begin{enumerate}

\item {\underline {QGP scenario}:} The system starts 
evolving from an initial QGP state formed at a time $\tau_i$ 
and temperature $T_i$.
It then expands, hence cools and reaches the critical 
temperature $T_c$  at a time $\tau_q$.
In the mixed phase the cooling due to expansion is
compensated by the heating of the system due
to the liberation of the latent heat in a first order
phase transition. Hence the temperature remains constant 
at $T_c$ (super-cooling is neglected here).
After the mixed phase ends at a 
time $\tau_{h}$, further expansion 
takes place and finally the system disassembles at a time $\tau_{f}$
and temperature $T_{f}$. At this stage, called the freeze-out stage,
the mean free path of the particles are too large to have any further interactions,
and those are detected experimentally.

\item {\underline {Hadron gas scenario}:} 
The hot and dense system is formed in the hadronic 
state at a time $\tau_i$ and temperature $T_{i}$ and
the system expands from the initial to freeze-out state
(at time $\tau_{f}$ when the temperature is  $T_{f}$) 
without a phase transition. 

\item {\underline {Hadron gas with mass variation}:}
The other possibility is that the system
may form in the hadronic state as in (2) above, but the
spectral function of the  hadrons  are different
from their vacuum counterparts. Among others, in the present case we will
consider the shift of the pole of the hadronic spectral
function according to the universal scaling law proposed
by Brown and Rho~\cite{br}:
\begin{equation}
m_h^* = m_h\left(1 - \frac{T^2}{T_c^2}\right)^\lambda,
\label{eq2}
\end{equation}
where $m_h^*$ ($m_h$) is the in-medium (vacuum) mass of the hadrons
(except pseudo-scalar)
and the index $\lambda$ takes a value between 0 and 1.
Here we choose $\lambda=1/6$ according to the well known Brown-Rho 
scaling~\cite{brpr}.
\end{enumerate}

\section{Initial Conditions}

The initial conditions in terms of initial temperature ($T_{i}$)
can be set for the 
three scenarios from the following relation:

\begin{equation}
dS = \frac{2 \pi^4}{45 \zeta(3)}\,dN 
              = 4 \frac{\pi^2}{90}\,g_{eff}\,T_{i}^{3}\,\Delta\,V,
\label{eq3}
\end{equation}
where $dS (dN)$ is the entropy (number) contained within a volume 
element $\Delta\,V=\pi\,R^2\,\tau_i\,d\eta$, $R$ as the radius of 
the colliding nuclei. $g_{eff}$ is the effective statistical degeneracy,
$\zeta(3)$ denotes the Riemann zeta function
and $\eta$ is the space-time rapidity. For massless  bosons
(fermions) the ratio of $dS$ to $dN$ is given by 
${2 \pi^4}/({45 \zeta(3)})\,\sim\,3.6$ (4.2), which is a crude
approximation for heavy particles. 
For example the above ratio is 3.6 (7.5) for 140 MeV pions (938 MeV protons)
at a temperature of 200 MeV.

First we consider the situation at SPS energies where $dN/dy$ $\sim$ 700 
for Pb+Pb collisions. For above three scenarios taking $g_{eff}$ as given in
Table~\ref{table1}, we obtain the initial temperature by using Eqn.~\ref{eq3}.
The values of the initial temperatures are given in Table~\ref{table1}.

The chemical potential at the initial state is fixed 
by constraining the specific entropy (entropy per baryon)
to the value obtained from the analysis of experimental data.
The specific entropy at SPS is about $40$ for Pb + Pb 
collisions~\cite{pbm,cleymans,roland,na49}. The net baryon number can
be calculated using the baryon (anti-baryon) number density, $n_b$ 
($n_{\bar b}$) given by,
\begin{equation}
n_{b}  = \frac{g}{(2 \pi)^3} \int{} f(\vec {p}) d^{3} p,
\label{eq4}
\end{equation}
where $g$ is the baryonic degeneracy. We take $g~=~4$ for 
proton and neutron.  $f(\vec {p})$ is
the well known Fermi-Dirac distribution, 
\begin{equation}
f(\vec {p}) = \Bigl[ { exp \Bigl( (E \pm \mu)/T \Bigr) + 1 } \Bigr]^{-1}
\label{eq5}
\end{equation}
where $\mu (-\mu) $ is the chemical potential for
baryon (anti-baryon),  $E = \sqrt{ p^2 + m^2}$ and
$n_{\bar b}\equiv\,n_b(\mu\rightarrow -\mu)$. 

For the QGP scenario we take the  mass of the quarks as,
${m_{q}}^{2} = {m_{qc}}^{2} + {m_{qth}}^{2}$, where 
$m_{qth}$ is the thermal mass~\cite{bellac} and $m_{qc}$ is the current
quark mass.  We have taken vacuum mass and effective mass 
for the hadrons (given by Eqn.~\ref{eq2})  for cases 2 and 3 respectively.
We obtain the value of the chemical potential to be
132 MeV, 340 MeV and 105 MeV for cases (1), (2) and
(3) respectively. These are also summarized in Table~\ref{table1}.

\begin{table}
\caption{Initial conditions and the initial values of fluctuation for the three scenarios. \\
\label{table1}}
\begin{tabular}{|lccc|}
\tableline
Initial Values/Scenarios        & 1 & 2 & 3\\
\tableline
$g_{eff}$       & 37  & 15  & 24  \\
&  &  & \\
$T_{i}$ (MeV)   & 196 & 264 & 226 \\
&  &  & \\
$\mu_{i}$ (MeV) & 132 & 340 & 105 \\
&  &  & \\
${(\Delta N_b (\tau_{i}))^2}/{S}$ & 0.014 & 0.029 & 0.061 \\
&  &  & \\
${(\Delta N_b (\tau_{i}))^2}/{N_{b,y}}$ & 0.56 & 1.16 & 2.46 \\
\tableline
\end{tabular}
\end{table}

Having fixed the initial temperature and chemical potential
we now calculate the initial fluctuations for the three scenarios. \\

\noindent {\it {1. Quark Gluon Plasma}:} 
From the initial net baryon number, one can easily calculate the
net baryon number fluctuations using Eqn.~\ref{eq1}.
The fluctuation in the net baryon number in the QGP
phase at time $\tau_{i}$, temperature $T_{i}$ and chemical potential $\mu_{i}$,
can be shown to be~\cite{twin_prl2}:
\begin{equation}
(\Delta N_b (\tau_{i}))_{\rm QGP}^2 = {2V\over 9 } T_{i}^3
\left(1 + {1\over 3} \Bigl({\mu_{i}\over \pi T_{i}}\Bigr)^2 \right) \,  ,
\label{eq6}
\end{equation}
With the initial conditions as discussed above,
the initial fluctuations for the QGP scenario turns out to be
$(\Delta N_b (\tau_{i}))_{\rm QGP}^2$ = 35.
It may be noted that the value of $g$ in Eqn.~\ref{eq4} is taken to be
12, for two flavor case.

The entropy density in the QGP phase is calculated from Eqn.~\ref{eq3}
with $g_{eff}$ = 37. The total entropy of the system is given by:
\begin{equation}
S = V \frac{4 \pi^{2} g_{eff}}{90 }T^{3}_{i} \sim 2500.
\label{eq7}
\end{equation}
For fixed initial entropy of the system the initial temperature
is different for the three cases because of the different values
of the $g_{eff}$. In the present work we solve the evolution 
equation for ideal fluid neglecting entropy generations
due to various viscous effects. 
The total entropy is kept constant for all the three scenarios as it is
obtained from the number of particles per unit rapidity 
measured experimentally.
So the fluctuation in net baryon number per entropy is 
${(\Delta N_b (\tau_{i}))_{\rm QGP}^2}/{S}$ = 0.014, a
value similar to that obtained in ~\cite{twin_prl2}.
In this case the fluctuation per unit baryon,
${(\Delta N_b (\tau_{i}))_{\rm QGP}^2}/{N_{b,y}}$ = 0.56
where $N_{b,y}$ is the net baryon number per unit 
rapidity, $dN_b/dy\sim 62$ for SPS energies. \\

\noindent {\it {2. Hadron Gas }:} 
The net baryon number fluctuation in the hadronic gas can be calculated using
Eqn.~\ref{eq1} and is given by:
\begin{equation}
(\Delta N_b)_{\rm HG}^2 =
{g V\over {4}} \Bigl[ {2 m T_{i}\over {\pi}} \Bigr]^{3/2}\, exp(-m/T_{i}) \cosh(\mu/T_{i})\, . 
\label{eq8}
\end{equation}
Substituting the values of $T_{i}$, $\mu_{i}$
and $m~=~938$ MeV for the nucleons, 
we obtain the fluctuation in net baryon number,
$(\Delta N_b (\tau_{i}))_{\rm HG}^2\,\sim\,72$. 
The ratio of the fluctuation to total entropy is 
${(\Delta N_b (\tau_{i}))_{\rm HG}^2}/{S}$ = 0.029.
This value is also 
similar to the one obtained in~\cite{twin_prl2}. In this case
the value of ${(\Delta N_b (\tau_{i}))_{\rm HG}^2}/{N_{b,y}}$ =1.16. \\

\noindent {\it {3. Hadron gas with mass variation in medium}:} 
The initial fluctuation for this case
can be obtained using the Eqn's~\ref{eq4} and ~\ref{eq5} 
together with Eqn.~\ref{eq2} used for the value of mass. The
initial conditions of $T_{i}$ = 226 MeV and $\mu_{i}$ = 105 MeV
are constrained to reproduce the hadronic multiplicity 
and the entropy per baryons. In this case we get
${(\Delta N_b (\tau_{i}))_{\rm HG,m^{*}}^2}$=153 and
${(\Delta N_b (\tau_{i}))_{\rm HG,m^{*}}^2}/{S}$ = 0.061.
The value of ${(\Delta N_b (\tau_{i}))_{\rm HG,m^*}^2}/{N_{b,y}}$ =2.46
here. \\

We observe that for the baryon number fluctuations corresponding
to the above three scenarios:
\begin{equation}
\frac {(\Delta N_b (\tau_{i}))_{\rm HG}^2}
{(\Delta N_b (\tau_{i}))_{\rm QGP}^2} \sim 2\,\,\,\, \newline
{\rm and} \newline
\frac {(\Delta N_b (\tau_{i}))_{\rm HG,m^*}^2}
{(\Delta N_b (\tau_{i}))_{\rm QGP}^2} \sim 4.\,\,\,\, \newline
\label{phased}
\end{equation}
This clearly shows that in the initial stage, there is a clear
distinction between the three cases. 
The initial values of fluctuations are summarized in Table~\ref{table1}.
It will be of interest
now to see, if these fluctuations (and the differences 
given by Eqn.~\ref{phased})  survive till the freeze-out.
Next we discuss the evolution of this initial fluctuation for
the three scenarios.

\section{Evolution of the initial fluctuation}
We follow ref.~\cite{twin_prl2} to study the 
proper time ($\tau$) evolution of the baryon fluctuation
in the space-time rapidity interval $\Delta\,\eta$ for
different EOS as well as initial states as mentioned 
above.  At the freeze-out point the fluctuation 
measured experimentally contain the residue of the
initial fluctuation which survived the space time 
evolution and the fluctuations generated due to the
exchange of baryons with the adjacent sub-volumes.
We discuss each of the above cases 
separately in the following sub-sections.

\subsection{Dissipation of the initial fluctuation}

The difference in baryon
flux ($\Delta\,n\,{\bar v}$) 
originating from different densities inside and outside of 
the sub-volume ($A\tau\Delta\,\eta$)
leads to the following differential equation for 
$\Delta{N_b} (=\Delta\,n{\bar v}A\tau)$:

\begin{equation}
{d{\Delta{N_b}}\over{d\tau}} = 
-{{\bar v} \over{2 \Delta{\eta}}} \frac{\Delta{N_b}}{\tau}
\label{eq9}
\end{equation}
where ${\bar v}$ is the average thermal  velocity 
of the particles under consideration.

The solution to the differential equation~\ref{eq9} 
from initial time $\tau_{i}$ to final time $\tau_{f}$ is then given as:
\begin{equation}
\Delta N_b(\tau_{f}) = \Delta N_b{(\tau_{i})} \exp\left(
-{1\over 2\Delta\eta} \int_{\tau_i}^{\tau_{f}} {d\tau\over\tau}\,
{\bar v}(\tau) \right) \, .
\label{eq10}
\end{equation}

\begin{figure}
\begin{center}
\vspace{-1cm}
\includegraphics[scale=0.5]{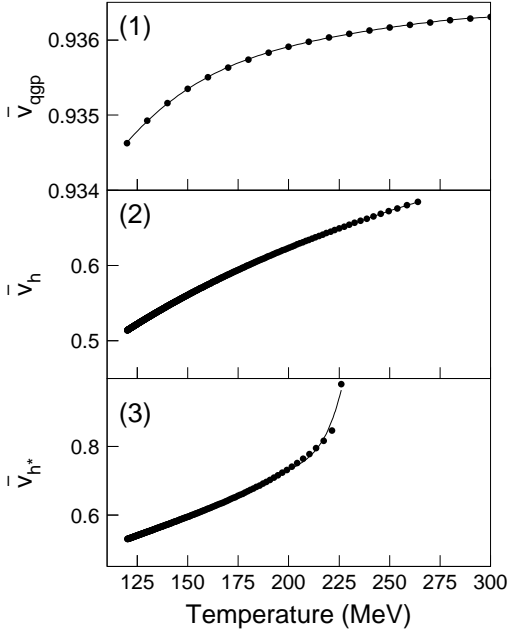}
\caption{ 
Variation of average thermal velocity with temperature for three different
scenarios. The results are fitted by  nth-order polynomials.
(1) For QGP scenario the fitting parameters are
$a_0~=~0.93$, 
$a_1~=~0.11$, $a_2~=~-0.56$, $a_3~=~1.32$ and $a_4~=~1.20$.
(2) For hadronic gas scenario with fit parameters as 
$a_0~=~0.19$, 
$a_1~=~4.0$, $a_2~=~-14.4$, $a_3~=~32.2$ and $a_4~=~-31.2$.
(3) For hadronic gas with mass variation in medium with fit parameters
as
$a_1~=~-4889.8$, $a_2~=~76879$, $a_3~=~-0.640E+06$, 
$a_4~=~0.297E+07$,
$a_5~=~-0.73E+07$ and $a_6~=~0.74E+07$.
}
\label{ave_velo}
\end{center}
\end{figure}

The average thermal velocity at a given temperature 
can be calculated by using the following equation,
\begin{equation}
{\bar v}  = \frac{
\int_{} {p\over E} {d^{3}p}\,
{f(\vec {p})}}{\int_{}{d^{3}p}{f(\vec {p})}} \,    \label{eq11}  
\end{equation}
The values of $\bar{v}$ as a function of temperature for the
three cases are shown in Fig. ~\ref{ave_velo}. 
The curves are fitted by the polynomials of the form
$a_{0} + a_{1}T + a_{2}T^{2} 
+ a_{3}T^{3} + a_{4}T^{4} + a_{5}T^{5} + a_{6}T^{6}$. 
The values of the parameters are 
shown in the caption of Fig.~\ref{ave_velo}.
The average velocity in QGP is found to be substantially larger
than the velocity in hadronic gas. However, in case of the hadronic 
system where the effective masses of the baryons (neutron and proton 
here) approach zero at $T_c$, their velocities approach the velocity of light.

\begin{figure}
\begin{center}
\includegraphics[scale=0.4]{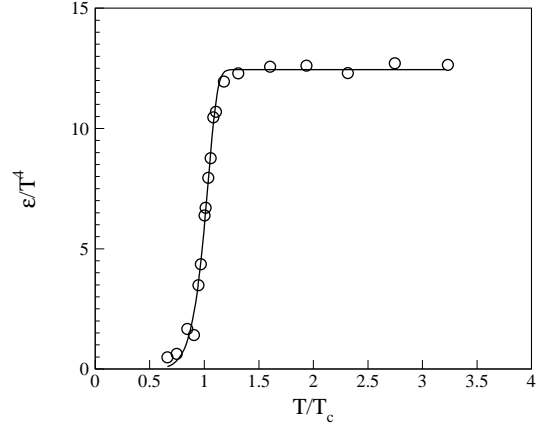}
\caption{ 
The variation of energy density as a function of temperature. 
The result shown by solid line is obtained from Eqn.~\protect{\ref{eq12}}.
Open circles show the lattice results with out the error bars.
}
\label{lat_par}
\end{center}
\end{figure}

The equation of state plays a vital role in deciding 
how the fluctuations in the net baryon number will evolve.
The variation of energy density ($\epsilon$)
as a function of temperature is obtained from the 
3-flavor lattice QCD results~\cite{karsch} which is
parametrized as follows:
\begin{equation}
\epsilon = T^{4} A~tanh( B (\frac{T}{T_{c}})^C ),
\label{eq12}
\end{equation}
where the values of the parameters, $A, B$ and $C$ are 12.44, 0.517
and 10.04 respectively. Note that the effect of net baryons in the EOS
is neglected here.
Fig.~\ref{lat_par} shows the variation of the energy density
with temperature. The increase in the effective degeneracy
near $T_c$ can be obtained from the hadronic phase 
with effective masses varying with temperature as in Eqn.~\ref{eq2}.
The increase in the effective degeneracy originates
from the heavier hadrons going to a massless 
situation~( see also ref.\cite{kb,bbjp}).

The evolution of the system under the boost invariance along
the longitudinal direction is governed by the equation~\cite{jdb},
\begin{equation}
\frac{d \epsilon}{d \tau}  + \frac { \epsilon + P}{\tau} 
= 0;\,\,\,\,P = c_{s}^2 \epsilon
\label{eq13}
\end{equation}
where $c_{s}$ is the velocity of the sound in the medium.
To perform the integration in Eqn.~\ref{eq10} we need $d\tau/\tau$, which
can 
be  obtained as

\begin{equation}
\frac{d\tau}{\tau} = - \Bigl[ { \frac{\beta}{T} + \frac{\alpha}{\chi} 
\frac{d\chi}{dT} } \Bigr] dT = - f(T) dT
\label{eq14}
\end{equation}
where, $\alpha$ = $1 \over{1+c_{s}^2}$, $\beta = 4\alpha$ and
\begin{equation}
\chi=\Bigl[tanhB(T/T_c)^C\Bigr]^\alpha
\end{equation}
The expression for the second term within the square bracket 
of Eqn.\ref{eq14} is given by
\begin{equation}
\frac{1}{\chi} \frac{d\chi}{dT} = \alpha \frac{BC}{T} (\frac{T}{T_c})^C 
\frac{1}{cosh(B (\frac{T}{T_c})^C)~ sinh(B (\frac{T}{T_c})^C)}
\label{eq15}
\end{equation}

Using the relation $c_s^2=\frac{S}{T}\frac{dS}{dT}=[3+\frac{T}{g_{eff}}\,\frac{dg_{eff}}{dT}]^{-1}$ 
we get the velocity of sound corresponding to the lattice QCD calculations,
\begin{equation}
c_{s}^2 = \Bigl[ 3 + BC (\frac{T}{T_c})^C \frac{1}{cosh(B (\frac{T}{T_c})^C)~ sinh(B (\frac{T}{T_c})^C)} \Bigr]^{-1}
\label{eq16}
\end{equation}

Now the fluctuation at the freeze-out point can be written as,
\begin{equation}
\Delta N_b(\tau_{f}) = \Delta N_b{(\tau_{i})} \exp\left(
-{1\over 2\Delta\eta} \int_{T_f}^{T_{i}} f(T) dT\,
{\bar v}(T) \right) \,.
\label{eq17}
\end{equation}

In order to study the effect of the equation of state, 
we will present results for three different values 
of $c_s$: (a) $c_s^2=1/3$ corresponds to the ideal gas case, (b) $c_s^2=0.18$  
corresponding to an EOS of hadronic gas where particles 
of mass upto 2.5 GeV has been taken into account  from particle 
data book and (c) $c_s^2$ obtained from Eqn.~\ref{eq16}. 
It may be mentioned that for the ideal gas the rate of cooling is faster
than the other two cases, (b) and (c).

Now we shall calculate the total dissipation of fluctuation at the freeze-out
temperature. We shall consider the freeze-out temperature to be 
120 MeV~\cite{na49,wa98} and a critical temperature for the QGP 
transition to be 170 MeV~\cite{karsch}. In all the three cases below we 
have taken the value of $\Delta \eta$ = 1. \\

\noindent {\it {1. Quark Gluon Plasma}:} 
For the QGP initial state, the dissipation equation has three parts.
In the first part, we calculate the dissipation from the 
initial temperature $T_{i}$ to $T_{c}$ (for the QGP phase).
The second part is the dissipation during the mixed phase and
in the final phase it is the dissipation from $T_{c}$ (end of mixed phase)
to the freeze-out temperature $T_{f}$. Hence the complete dissipation
equation becomes,

\begin{widetext}
\begin{eqnarray}
\Delta N_b(T_{f}) = \Delta N_b{(T_{i})} \exp\left(
-{1\over 2\Delta\eta} \left( \int_{T_{c}}^{T_{i}}\, 
{\bar v_{qgp}}(T) f(T) {dT}  + {\bar v_{mix}}(T_{c})~ln(r)~ 
+ \int_{T_{f}}^{T_{c}} {\bar v_{h}}(T) f(T) {dT} \right) \right)
\label{eq18}
\end{eqnarray}
\end{widetext}
where $r$ is the ratio ${(g_{eff})_{qgp}}/{g_{eff})_h}$ $\sim$ 2.5
and ${\bar v_{mix}}\,\sim\,{\bar v_{qgp} (T_{c})}$.

The fluctuation in the net baryon number is evaluated for:
(a) $c_{s}^2$ = 1/3, for which we get
$(\Delta N_b (\tau_{f}))_{\rm QGP}^2 \sim 6$,
(b) $c_{s}^2$ = 0.18, giving
$(\Delta N_b (\tau_{f}))_{\rm QGP}^2 \sim 2.23$ and 
(c) $c_{s}^2$ as given by Eqn.~\ref{eq16}, resulting in
$(\Delta N_b (\tau_{f}))_{\rm QGP}^2\,\sim  0.54$. 

\begin{figure}
\begin{center}
\vspace{-1.4cm}
\includegraphics[scale=0.5]{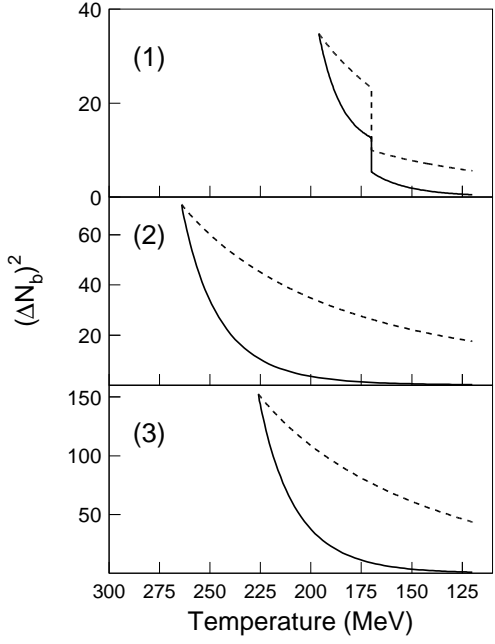}
\caption{ 
The dissipation of net baryon number fluctuation  for the different scenarios
as a function of temperature. 
Note that temperature is plotted in decreasing
order to reflect the evolution in time.
(1) Variation of ${(\Delta N_b)^2}$ for QGP scenario,   
(2) Variation of ${(\Delta N_b)^2}$ for hadronic gas scenario  and
(3) Variation of ${(\Delta N_b)^2}$ for hadronic gas with mass 
variation scenario.
The dashed lines correspond to results obtained for $c_{s}^2$ = 1/3, while the
solid lines show the results obtained using the value of $c_{s}^2$ 
given in Eqn.~\ref{eq16}. 
}
\label{evovl_fluc_sps}
\end{center}
\end{figure}

The initial and final fluctuation values along with the evolution process
of the values for the QGP scenario is shown in Fig.~\ref{evovl_fluc_sps}.
Only the results for  $c_{s}^2$ = 1/3 and $c_{s}^2$ as given by Eqn.~\ref{eq16}
are shown in the figures for the clarity of presentation. 
We have checked that the results for 
$c_{s}^2$ = 0.18 lies between the above two cases. \\

\noindent {\it {2. Hadron Gas }:} 
In case of hadronic initial state we have,
\begin{eqnarray}
\Delta N_b(T_{f}) = \Delta N_b{(T_{i})} \exp\left(
-{1\over 2\Delta\eta}  \int_{T_{f}}^{T_{i}} {\bar v_{h}}(T) f(T) {dT} \right)
\label{eq19}
\end{eqnarray}
(a) For $c_{s}^2$ = 1/3, the contribution from the exponential
is $\sim$ 0.5. Hence the fluctuation in net baryon number is  
$(\Delta N_b (\tau_{f}))_{\rm HG}^2$ is $18$,
(b) for $c_{s}^2$ = 0.18,  we get
$(\Delta N_b (\tau_{f}))_{\rm HG}^2$ is $14.6$ and 
if we take  (c) $c_{s}^2$ from Eqn.~\ref{eq16} the contribution from   
the exponential term is $\sim$ 0.06, a factor of 8 
smaller than the value for case (a). Consequently
the fluctuation in net baryon number is  
$(\Delta N_b (\tau_{f}))_{\rm HG}^2$ is $0.22$. 
The evolution of fluctuation for this case is depicted in 
Fig.~\ref{evovl_fluc_sps}. \\

\begin{figure}
\begin{center}

\includegraphics[scale=0.5]{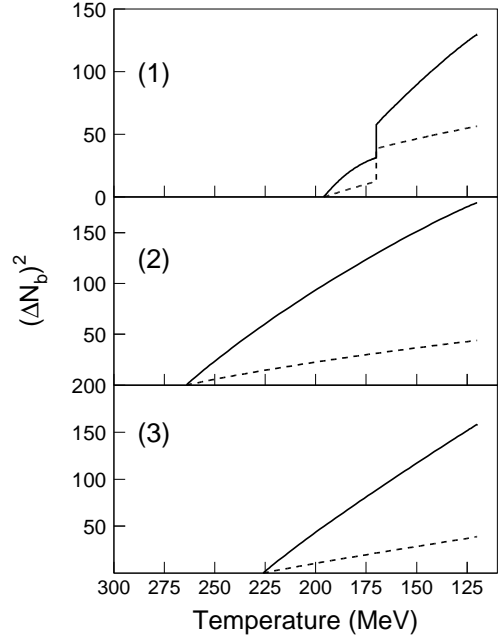}
\caption{
The generation of net baryon number fluctuation  for the different scenarios
as a function of temperature. Notations are same as that of 
Fig.~\protect{\ref{evovl_fluc_sps}}.
}
\label{evovl_gen_sps}
\end{center}
\end{figure}

\noindent {\it {3. Hadron gas with mass variation in medium}:} 
For the hadronic initial state  with mass variation,
\begin{eqnarray}
\Delta N_b(T_{f}) = \Delta N_b{(T_{i})} \exp\left(
-{1\over 2\Delta\eta}\int_{T_{f}}^{T_{i}} {\bar v^{*}_{h}}(T) f(T) {dT} \right)
\label{eq20}
\end{eqnarray}
(a) For $c_{s}^2$ = 1/3, the contribution from the exponential  
is $\sim$ 0.53. Hence the fluctuation in net baryon number is  
$(\Delta N_b (\tau_{f}))_{\rm m^{*}}^2 \sim 44$,
(b) for $c_{s}^2$ = 0.18, we get:
$(\Delta N_b (\tau_{f}))_{\rm m^{*}}^2 \sim 37$ and
(c) taking $c_{s}^2$ from Eqn.~\ref{eq16}, the contribution 
from the exponential term is $\sim$ 0.08 a  factor of 7 lower
than the case (a). So the fluctuation in net baryon number is  
$(\Delta N_b (\tau_{f}))_{\rm HG, m^{*}}^2$ is $0.92$.
The evolution of fluctuation for this case is shown in 
Fig.~\ref{evovl_fluc_sps}. \\

\subsection{Generation of fluctuation with time}
The baryon fluxes exchanged with neighboring sub-volumes can lead to
generation of fluctuations, and is the main source that is to be
detected experimentally at the freeze-out point.
The total number of baryons leaving or entering
($N_{b}^{ex}$) the sub-volume ($A \tau \Delta \eta$) 
between time $\tau_{i}$ and $\tau_{f}$ is given by~\cite{twin_prl2}
\begin{eqnarray}
N_b^{ex}(T_{f}) =  {N_b{(T_{i})}\over 2\Delta\eta} 
\left( \int_{T_{f}}^{T_{i}}\, {\bar v}(T) f(T) {dT} \right) 
\label{eq21}
\end{eqnarray}

\noindent {\it {1. Quark Gluon Plasma}:} 
As in the previous case the fluctuations has to be evaluated in three 
parts in a QGP formation scenario.
In the first part, we calculate the generation from the 
initial temperature $T_{i}$ to $T_{c}$ (for the QGP phase).
The second part is the generation during the mixed phase and
in the final phase it is the generation from $T_{c}$ (end of mixed phase)
to the freeze-out temperature $T_{f}$. Hence the complete evolution
equation becomes,

\begin{widetext}
\begin{eqnarray}
(\Delta N_b(T_{f}))^2 = N_b(T_i)\left(
{1\over 2\Delta\eta} \left( \int_{T_{c}}^{T_{i}}\, 
{\bar v_{qgp}}(T) f(T) {dT}  + {\bar v_{mix}}(T_{c})~ln(r)~ 
+ \int_{T_{f}}^{T_{c}} {\bar v_{h}}(T) f(T) {dT} \right) \right)
\label{eq22}
\end{eqnarray}
\end{widetext}

The fluctuation in the net baryon number is evaluated for:
(a) $c_{s}^2$ = 1/3, for which we get
$(\Delta N_b (\tau_{f}))_{\rm QGP}^2$ is $56.5$, 
(b) $c_{s}^2$ = 0.18,  resulting in
$(\Delta N_b (\tau_{f}))_{\rm QGP}^2$ is $60.4$ and
(c) $c_{s}^2$ from Eqn.~\ref{eq16} giving
$(\Delta N_b (\tau_{f}))_{\rm QGP}^2$ is $129$. 

The initial and final values of the fluctuations along with its 
time evolution for the QGP scenario are shown in Fig.~\ref{evovl_gen_sps}.

\noindent {\it {2. Hadron Gas }:} 
In case of hadronic initial state we have,
\begin{eqnarray}
(N_b(T_{f}))^2 =N_b(T_i)\left(
{1\over 2\Delta\eta}  \int_{T_{f}}^{T_{i}} {\bar v_{h}}(T) f(T) {dT} \right)
\label{eq23}
\end{eqnarray}
(a) For $c_{s}^2$ = 1/3, the fluctuation in net baryon number is  
$(\Delta N_b (\tau_{f}))_{\rm HG}^2$ is $44$. 
(b) For $c_{s}^2$ = 0.18, we get
$(\Delta N_b (\tau_{f}))_{\rm HG}^2$ is $50$. 
(c) If we take  $c_{s}^2$ from Eqn.~\ref{eq16}, 
the fluctuation in net baryon number is  
$(\Delta N_b (\tau_{f}))_{\rm HG}^2$ is $179$. 
The evolution of fluctuation for this case is displayed in 
Fig.~\ref{evovl_gen_sps}.  \\

\noindent {\it {3. Hadron gas with mass variation in medium}:} 
For the hadronic initial state  with mass variation,
\begin{eqnarray}
(\Delta N_b(T_{f}))^2 =N_b(T_i)\left(
{1\over 2\Delta\eta}  \int_{T_{f}}^{T_{i}} {\bar v^{*}_{h}}(T) f(T) {dT} \right)
\label{eq24}
\end{eqnarray}
(a) For $c_{s}^2$ = 1/3, the fluctuation in net baryon number is  
$(\Delta N_b (T_{f}))_{\rm m^{*}}^2 \sim 38.9$.
(b) For $c_{s}^2$ = 0.18, we get
$(\Delta N_b (T_{f}))_{\rm m^{*}}^2 \sim 44$.
(c) Taking $c_{s}^2$ from Eqn.~\ref{eq16}, 
the fluctuation in net baryon number is  
$(\Delta N_b (T_{f}))_{\rm HG, m^{*}}^2$ is $158$.
The evolution of fluctuation for this case is shown in 
Fig.~\ref{evovl_gen_sps}.

\section{Fluctuations at the freeze-out}

The net baryon fluctuation at the freeze-out ($T_f$ = 120 MeV) 
is a combination of the
dissipation and the generation effect as presented in the previous section.
The resultant fluctuation is the sum of the variances $(\Delta N_b(T_{f}))^2$
obtained for each of the two processes. 

We present results in terms of net baryon fluctuation 
per unit baryon, 
${(\Delta N_b (T_{f}))^2}/{N_{b,y}}$. For Poissonian noise this
value should be close to unity. Deviation  
from this numerical values will indicate the presence of 
dynamical fluctuation.

\noindent {\it {1. Quark Gluon Plasma}:} 
The final fluctuation in the net baryon number per unit baryon in the QGP
scenario at the freeze-out point for the  three EOS mentioned above
are given by,
(a) for $c_s^2$ = 1/3,
${(\Delta N_b (T_{f}))_{\rm QGP}^2}/{N_{b,y}}$ $\sim$ 1.0,
(b) for $c_s^2$ = 0.18,
${(\Delta N_b (T_{f}))_{\rm QGP}^2}/{N_{b,y}}$ $\sim$ 1.0 and
(c) taking $c_{s}^2$ from Eqn.~\ref{eq16}, 
${(\Delta N_b (T_{f}))_{\rm QGP}^2}/{N_{b,y}}$ $\sim$ 2.0. \\

\noindent {\it {2. Hadron Gas }:}  
The net baryon number fluctuation per unit baryon 
in the hadronic gas at freeze-out for different values of
of $c_s$ are,
(a) for $c_s^2$ = 1/3,
${(\Delta N_b (T_{f}))_{\rm HG}^2}/{N_{b,y}}$ $\sim$ 1.0
(b) for $c_s^2$ = 0.18,
${(\Delta N_b (T_{f}))_{\rm HG}^2}/{N_{b,y}}$ $\sim$ 1.0
(c) Taking $c_{s}^2$ from Eqn.~\ref{eq16}, 
${(\Delta N_b (T_{f}))_{\rm HG}^2}/{N_{b,y}}$ $\sim$ 2.8. \\

\noindent {\it {3. Hadron gas with mass variation in medium}:} 
The net baryon number fluctuation per unit baryon 
for the hadronic initial state with mass variation 
at freeze-out for the three EOS are,
(a) for $c_s^2$ = 1/3,
${(\Delta N_b (T_{f}))_{\rm HG,m^*}^2}/{N_{b,y}}$ $\sim$ 1.3,
(b) for $c_s$ = 0.18,
${(\Delta N_b (T_{f}))_{\rm HG,m^*}^2}/{N_{b,y}}$ $\sim$ 1.3 and
(c) taking $c_{s}^2$ from Eqn.~\ref{eq16}, 
${(\Delta N_b (\tau_{f}))_{\rm HG,m^*}^2}/{N_{b,y}}$ $\sim$ 2.5. \\

The net baryon number fluctuations per unit baryon is summarized
in Table~\ref{table2}. For all the scenarios the values of the quantity 
$(\Delta N_b(\tau_f))^2/N_{b,y}$ is larger than Poissonian 
noise for EOS taken from lattice QCD. This indicates presence of 
dynamical fluctuations.
However, it should be mentioned here that the
errors in lattice QCD calculations is large for temperature below the critical
temperature. Furthermore 
$(\Delta N_b (\tau_f))_{\rm HG}^2/
(\Delta N_b (\tau_f))_{\rm QGP}^2 \sim 1.4$\,\,
and
$(\Delta N_b (\tau_f))_{\rm HG,m^*}^2/
(\Delta N_b (\tau_f))_{\rm QGP}^2 \sim 1.25$.
This means, although scenario (1) may be distinguishable
from the scenarios (2) and (3) it is very difficult
to differentiate  (2) and (3) from fluctuation  
measurements. However, for EOS with $c_s^2=1/3$ and $0.18$
they are close to Poisson value of 1. This indicates that
for these EOS; it is difficult to distinguish among the three scenarios.

\begin{table}
\caption{$(\Delta N_b)^2/N_{b,y}$ at freeze-out for three different evolution
scenarios at SPS and RHIC energies. \\
\label{table2}}
\begin{tabular}{|clccc|}
\tableline
     & EOS/Scenarios                   & 1 & 2 & 3\\
\tableline
     & $c_{s}^{2}$ = 1/3  & ~~~1.0~~~  & ~~~1.0~~~  & ~~~1.3~~~  \\
     &                    &            &            &            \\
SPS  & $c_{s}^{2}$ = 0.18 & 1.0  & 1.0  & 1.3  \\
     &                    &            &            &            \\
     & Lattice            & 2.0  & 2.8  & 2.5  \\ 
     &                    &            &            &            \\
\hline
     &                    &            &            &            \\
     & $c_{s}^{2}$ = 1/3  & 1.46  & $-$  & $-$  \\
     &                    &            &            &            \\
~~RHIC~~ & $c_{s}^{2}$ = 0.18 & 1.96  & $-$  & $-$  \\
     &                    &            &            &            \\
     & Lattice            & 3.17  & $-$  & $-$  \\
\tableline
\end{tabular}
\end{table}

\section{Results at RHIC energies}

The net baryon number fluctuation has been evaluated for RHIC energies
of $\sqrt{s}=200 A\dot{\rm GeV}$ of Au+Au collisions.
We have taken the total (charged and neutral)
$dN/dy$ $\sim$ 1100 ~\cite{phobos}.  
The initial temperatures obtained from the above multiplicity are quite large, 
370 MeV for hadronic gas and 290 MeV for the case of hadronic 
gas with mass variation. As these temperatures are
well above the critical temperature predicted by lattice
QCD, we have considered only the QGP scenario for
RHIC energies.
We have taken the initial time to be $0.6$ fm/c~\cite{rhictime}
and specific entropy to be $150$ for Au + Au collisions.
For QGP we take $g_{eff}~=~47.5$ 
which gives $T_i\,\sim 251$ MeV and the 
constraint on the specific entropy gives an initial chemical potential of 
73 MeV. 
With these initial conditions for QGP scenario 
at RHIC, the initial fluctuations turn out to be : 
$(\Delta N_b (\tau_{i}))_{\rm QGP}^2$ = 42.
The entropy density in the QGP phase is calculated from Eqn.~\ref{eq7}
with $g_{eff}$ = 47.5 to be  $\sim$ 3900.
So the fluctuation in net baryon number per entropy is 
${(\Delta N_b (\tau_{i}))_{\rm QGP}^2}/{S}$ = 0.011 and
${(\Delta N_b (\tau_{i}))_{\rm QGP}^2}/{N_{b,y}}$ = 1.6.
The value of $dN_b/dy\sim 26$ for RHIC energies a factor of about 2.4 lower
than that for SPS energies.

\begin{figure}
\begin{center}
\vspace{-0.9cm}
\includegraphics[scale=0.4]{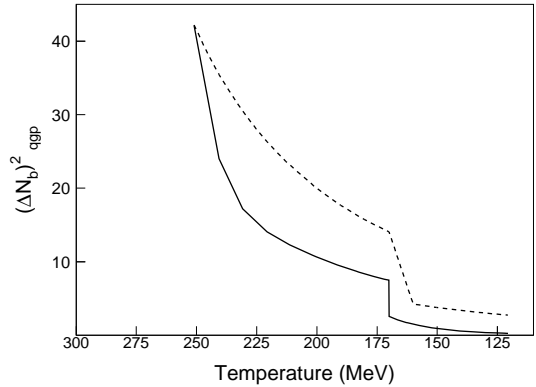}
\caption{ 
The dissipation of net baryon number fluctuation  for the QGP scenarios
as a function of temperature calculated for 
Au + Au collisions at RHIC energies. 
The dashed lines corresponds to results obtained for $c_{s}^2$ = 1/3, while the
solid lines corresponds to results obtained using the value of $c_{s}^2$ 
given in Eqn.~\ref{eq16}. 
}
\label{evovl_fluc_rhic}
\end{center}
\end{figure}

\begin{figure}
\begin{center}
\vspace{-0.9cm}
\includegraphics[scale=0.4]{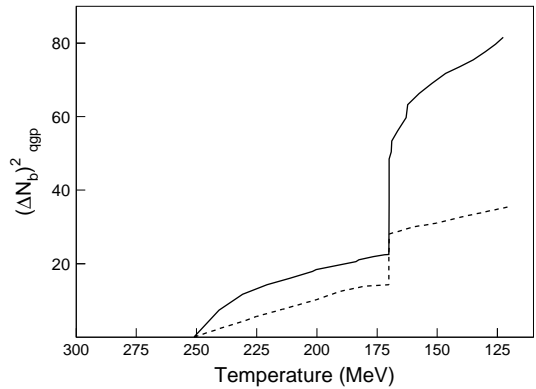}
\caption{ 
The generation of net baryon number fluctuation  for the QGP scenarios
as a function of temperature calculated at RHIC energies of 200 AGev
Au + Au Collisions. Notations are same as Fig.~\protect{\ref{evovl_fluc_rhic}}.
}
\label{evovl_gen_rhic}
\end{center}
\end{figure}

As before, here the values of $T_c$ and $T_f$ are 170 and 120 MeV
respectively.
(a) For $c_{s}^2$ = 1/3, the fluctuation in net baryon number 
due to dissipation at freeze-out, 
$(\Delta N_b (\tau_{f}))_{\rm QGP}^2$ turns out to be $2.7$.
Whereas the generation mechanism gives 
$(\Delta N_b (\tau_{f}))_{\rm QGP}^2\,\sim\, 35.5$.
So the resultant fluctuation in net baryon number per unit baryon is 
${(\Delta N_b (\tau_{f}))_{\rm QGP}^2}/{N_{b,y}}~\sim~1.46$.
(b) For $c_{s}^2$ = 0.18, the fluctuation in net baryon number 
due to dissipation, 
$(\Delta N_b (\tau_{f}))_{\rm QGP}^2$ turns out to be $2.3$.
Whereas the generation mechanism gives 
$(\Delta N_b (\tau_{f}))_{\rm QGP}^2\,\sim\, 48.7$.
The evolution of the dissipation and the generation of fluctuation
for the QGP scenario at RHIC energies is shown in the 
Figs.~\ref{evovl_fluc_rhic} and ~\ref{evovl_gen_rhic}. 
So the resultant fluctuation in net baryon number per unit baryon is 
${(\Delta N_b (\tau_{f}))_{\rm QGP}^2}/{N_{b,y}}~\sim~1.96$.
(c) If we take  $c_{s}^2$ as given by Eqn.~\ref{eq16},  
the fluctuation in net baryon number due to dissipation,   
$(\Delta N_b (\tau_{f}))_{\rm QGP}^2$ is $0.25$,
the generation mechanism gives 
$(\Delta N_b (\tau_{f}))_{\rm QGP}^2\,\sim\,82.25$.
So the resultant fluctuation in net baryon number per unit baryon is 
${(\Delta N_b (\tau_{f}))_{\rm QGP}^2}/{N_{b,y}}~\sim~3.17$.

Note that
though the absolute value of the fluctuations in case of RHIC 
is smaller than SPS the fluctuation per baryon at RHIC is larger
at the freeze-out point.

\section{Summary}

We have discussed the evolution of the
fluctuation in net baryon number
from the initial state to the final freeze-out 
state for three different scenarios, $viz.$,
(1) formation of QGP (2) hadronic gas and (3) hadronic gas with modified mass 
in the medium. In case 
of QGP formation we have assumed a first order phase transition.
We find that the fluctuations at the
initial stage agree with previously obtained values ~\cite{twin_prl2} where
there are clear distinctions among the three cases. The 
fluctuations at the freeze-out point depend crucially on 
the equation of state (value of $c_s$).
Fluctuations with ideal EOS is seen to dissipate at a 
slower rate compared to EOS from lattice calculation. 
At SPS energies the values of the variance depend crucially
on the EOS. For EOS from lattice QCD parametrization the 
fluctuation at the freeze-out is larger than the Poissonian noise.
However, for ideal gas EOS and EOS with $c_{s}^2$ = 0.18, 
we do not observe fluctuations of dynamical origin.
At RHIC energies the value of the fluctuations are larger
than the Poissonian noise for all the three EOS under consideration here,
indicating the fluctuations of dynamical origin.
The effects of the finite acceptance on the fluctuation enters
our calculations through $\Delta\eta$ and 
according to Eqn.~\ref{eq10} and~\ref{eq21} it is same for 
all the three scenarios (1), (2) and (3) discussed above.
For general discussions on the effects of 
acceptance on the fluctuations we refer to Ref.~\cite{twin_prl1,ijmpa}. 
The dependence of the fluctuation on  the centrality 
of the collisions (impact parameter) get canceled to a
large extent in the ratio,
${(\Delta N_b (\tau_{f}))_{\rm QGP}^2}/{N_{b,y}}$.
It is shown in ~\cite{wa98_fluc} that it is
possible to control the impact parameter dependence of
the fluctuation by measuring $E_T$ and analyzing data in narrow bins of 
$E_T$.  A full (3+1) dimensional 
expansion will lead to faster cooling, and hence it is interesting to see the 
survivability of fluctuations in such a scenario ~\cite{bm}.

\acknowledgments{One of us (B.M.) is grateful to the Board of Research
on Nuclear Science and Department of Atomic Energy, 
Government. of India for financial support. We would like to thank
M. Asakawa for useful comments. We are thankful to the referee for his
useful comments on the present manuscript.
}

\end{document}